\documentstyle[tighten,preprint,pra,aps]{revtex}
\begin{document}
\draft
\title{ Comments on ``Effective Core Potentials''
   \protect[ M.~Dolg, {\it Modern Methods and Algorithms of Quantum Chemistry}
    {\rm (Ed.\ by J.Grotendorst, John von Neumann Institute for Computing,
     J{\"u}lich, NIC Series, Vol.1, ISBN 3-00-005618-1, pp.479-508, 2000)}
   \protect].}
\author{A.~V.~Titov~\cite{E-mail} and N.~S.~Mosyagin}
\address{Petersburg Nuclear Physics Institute, \\
         Gatchina, Petersburg district 188350, Russia}
\date{\today}
\maketitle

\begin{abstract}
 The recent paper of M.~Dolg is discussed and his critical remarks with
 respect to the Generalized Relativistic Effective Core Potential (GRECP)
 method are shown to be incorrect. Some main features of GRECP are discussed as
 compared with the ``energy-consistent/adjusted'' pseudopotential and with the
 conventional shape-consistent RECP.
\end{abstract}

\section{Introduction}

 The discussed paper of M.\ Dolg deals with the relativistic effective core
 potential (RECP) methods including the model potential and pseudopotential
 (PP) techniques. The shape-consistent RECP method as a PP version is compared
 with the ``energy-adjusted'' PP (EAPP) and the ``energy-consistent'' PP (ECPP)
 developed by Stuttgart's group (e.g., see\ \cite{Kuchle,Hausser} and references
 in the discussed paper by M.~Dolg). In their semiempirical version, EAPP,
 partial potentials are fitted to reproduce the experimental atomic spectrum.
 In the ab initio approximation, ECPP (earlier also called energy-adjusted PP),
 ``valence energies'' (i.e.\ sums of ionization potentials and excitation
 energies) for a group of low-lying states are fitted to the corresponding
 energies of the same states in all-electron approximations like Hartree-Fock,
 Wood-Boring or Dirac-Fock in a least-squares sense with the help of the ECPP
 parameters. It means that only some special combinations of the matrix
 elements of a (non-,quasi-)relativistic Hamiltonian are fitted by the ECPP and
 EAPP Hamiltonians with the help of the one-electron radially local PP
 operator. When considering below both ECPP and EAPP versions we will write
 ECAPPs or Stuttgart PPs.

 The radially local operator is also used in the shape-consistent RECPs and
 some new non-local RECP terms are added in our Generalized RECP (GRECP)
 version
 \cite{tav91,Mos94,SfCRECP,Tup95,Mos97,Comm97,Mos98Hg,GRThGr,Mos00Hg,tav00GR,tav00TlH,tim00,Mos00HgH,tav00Ab},
 which we consider as a development of the shape-consistent RECP method.  The
 underlying idea traced in our papers concerning the GRECP approximation is in
 simulating the one- and two-electron parts of an original Dirac-Coulomb (in
 prospect, Dirac-Coulomb-Breit) Hamiltonian with the accuracy which is {\it
 needed and sufficient} for calculation of physical and chemical properties
 (and processes) in heavy-atom molecules with a {\it given} accuracy. A
 paramount requirement is that such a simulation should provide maximum
 possible savings for consequent molecular calculations with GRECPs.

 As is confirmed in all our test calculations and not only in our theoretical
 analysis, the GRECP Hamiltonian in the form used in papers by Mosyagin et al.\
 (1997)\ \cite{Mos97} and (2000) \cite{Mos00Hg} (which are criticized by
 M.~Dolg) more accurately reproduces the Dirac-Coulomb Hamiltonian {\it in the
 valence (V) region} as compared with other tested RECP and Stuttgart PP
 versions employing the radially local operator. Phrase ``in the valence
 region'' means that the occupation numbers of the outer core (OC) shells,
 $n_{OC}^{occ}$, are not considerably changed in studied states as compared
 with the OC occupation numbers of the configurations used in the GRECP
 generation (i.e.\ $\Delta n_{OC}^{occ} {\ll} 1$). Thus, only relaxation and
 dynamic correlation effects are suggested to take place in the OC shells.

 We have emphasized this property of the GRECP Hamiltonian already in the
 introduction of paper \cite{Mos97}. It is noted in the abstract of our
 theoretical paper (1997,1999)\ \cite{GRThGr} where a very detailed analysis of
 features of the shape-consistent RECP method including the GRECP approximation
 is given. We consider the GRECP version for Hg used in\ \cite{Mos97,Mos00Hg}
 as reliable for atomic and molecular calculations of the states in which the
 OC shells of Hg are completely occupied in the leading configurations if
 accuracy of a few hundreds wave numbers for transition, dissociation etc.\
 energies is required.
 In papers \cite{SfCRECP,GRThGr}, some other improvements of the RECP method
 are suggested in order to provide minimal computational efforts in accurate
 RECP calculation of wide range of excitations and properties in systems
 containing arbitrary heavy atoms including transition metals, lanthanides and
 actinides. We, obviously, will not repeat here the theoretical analysis of the
 shape-consistent RECP method and will give only some necessary details which
 have direct attitude to the criticized points.

 The goal of this paper is mainly to compare features of the GRECP and other
 RECP versions including Stuttgart PPs rather than to reply on Dolg's claims.
 These comments can be also useful for reading them before paper\ \cite{GRThGr}.

\section{General comparison of different RECP versions}
 \label{Comparison}

 The discussed GRECP version is assumed to be efficiently used when
 excitations and chemical bonding take place in the V region whereas only
 dynamic correlation and relaxation (polarization) are considered in the OC
 region. Therefore, interactions between/with valence and outer core electrons
 are simulated on the basis of the following principles:

 {\it First}, for selected subspaces of the OC and V shells, the matching
 radii $R_c$ for the regions of the spinor's smoothing are chosen to be as
 small as possible in order to reduce the errors of reproducing the original
 two-electron integrals (in further reducing the matching radii, partial
 potentials become too singular to be approximated by gaussians and used in
 RECP calculations, for details see\ \cite{GRThGr}).
 When using the GRECP operator in calculations of the same states as with the
 Dirac-Coulomb Hamiltonian, the OC, V and virtual pseudospinors coincide with
 the large components of the original Dirac spinors after the matching radii
 with significantly higher accuracy than in the cases of using the conventional
 RECP operator\ \cite{Tup95,GRThGr}. In a result (it is very important), the
 radius of the ``unphysical'' GRECP terms (the partial potentials $U_{nlj}(r)$
 contain the contribution from the Coulomb and exchange interaction with the IC
 electrons, the contribution from relativistic effects, and the contribution
 from {\it smoothing} the original spinors) only slightly larger than the
 outermost matching radius $R_c^{max}$ for pseudospinors. This is direct
 consequence of generating different partial potentials, $V_{nlj}(r)$, for the
 corresponding OC and V (and virtual) pseudospinors with the same ($lj$).  It
 is shown in\ \cite{tav00Ab} that difference between the OC and V potentials
 with the same ($lj$) can not be eliminated with the help of any special kind
 of smoothing the corresponding OC and V spinors without substantial decrease
 in accuracy for transition energies, up to an order of magnitude.
 Thus, when reducing the matching radii, which are usually close to
 each other, we reduce the radius of unphysical terms in the corresponding
 atomic effective Hamiltonian. 
 The independent smoothing the OC and V (and, in principle, some virtual)
 spinors with polinomials give us sufficient flexibility to generate smooth
 enough OC and V pseudospinors as well as their partial potentials.

 The GRECP is a ``matching radii-specified'' (or ``space-driven'') method of
 approximation (see\ \cite{GRThGr}) contrary to the energy-adjusted/consistent
 PPs which are ``selected valence energy-based'' (or ``energy-driven''). 
 The ``space-driven'' RECPs are substantially better justified from the
 theoretical point of view\ \cite{GRThGr} and, in general, they more properly
 reflect the relaxation/polarization properties of the inner core region in a
 heavy element.

 {\it Second}, the non-local terms with projectors on the outer core
 pseudospinors in the GRECP operator give us an important possibility to
 reproduce the original OC and V orbital energies and the (most important)
 nondiagonal Lagrange multipliers \cite{tav00Ab}, which together with the
 Hartree-Fock (Coulomb and exchange) electronic terms derived on pseudospinor
 densities constitute the one-electron part of the effective (model)
 Hamiltonian. Therefore, the one-electron part of the Hamiltonian in the V
 region can be reproduced very accurately (see section ``Theory'' in\
 \cite{GRThGr} for more details).
 The orbital energies are used in denominators of the M{\"o}ller-Plessett
 perturbation theory (PT) accounting for correlation etc.
 Otherwise, some other combinations of matrix elements of an original
 Hamiltonian can be reproduced in the model one instead of the orbital energies
 and nondiagonal Lagrange multipliers. (In particular, those one-electron
 energies can be exactly simulated which are more appropriate, e.g., for the
 Epstein-Nesbet PT. However, the partial potentials have additional ``tails''
 in such cases because of the use the inverted Hartree-Fock equations for their
 generation and the radius of the unphysical (G)RECP terms is thus enlarged.
 One should remember that the tail behaviour of orbitals is determined
 by their orbital energies.
 The distinction in using differently determined 
 one-electron energies is not very essential
 if the corresponding original Coulomb and exchange two-electron integrals are
 accurately reproduced by those with pseudospinors as is for small matching
 radii\ \cite{GRThGr}.)

 We are ``fitting'' the Hamiltonian matrix elements in the valence region
 first of all and not some their combinations of a special kind likewise the
 valence energies. It is sufficient to use a very small number of DF
 configurations for the GRECP generation, with the basic requirements:\
 {\it a)} they should have the same configurational structure in the core
 region as the states of the ``atom in a molecule'' studied in the GRECP
 calculations;\
 {\it b)} they should contain all the spinors required for the generation of
 the corresponding partial potentials. 

 Obviously, it is possible to fit transition or valence energies for a group
 of states with the help of the energy-consistent/adjusted PPs having
 appropriate number of fitting parameters. Certainly, it is not equivalent to
 the simulation of all the important Hamiltonian matrix elements on a {\it
 needed} level of accuracy in order to describe a possible variety of
 perturbations in the valence region of a considered element with a {\it given}
 accuracy. All the one- and two-electron matrix elements of the original
 valence Hamiltonian should be appropriately reproduced for the element to be
 used for accurate calculations of a wide range of applications including
 excitations and chemical bonding with arbitrary atoms and geometries. Besides,
 fitting the valence energies prior to the orbital energies, give no any
 advantages, in particular, in reproducing the variety of physical and chemical
 properties which cannot be calculated from potential curves or surfaces. The
 ``steady'' simulation of the valence Hamiltonian can be done on the basis of
 the ``space-driven'' shape-consistent RECP generation scheme.  

\vspace{0.3cm}
 The conventional RECP operator with the shape-consistent spinor smoothing
 suggested by K.Pitzer's group (1979,1977)\ \cite{Christ,Lee} gives no
 flexibility in fitting shapes and orbital energies simultaneously for the OC
 and V spinors with the same ($lj$). When the OC (pseudo)spinors are used in
 their RECP generation scheme for a given ($lj$), the ``effective matching
 radii'' for other (V and virtual) pseudospinors with the same ($lj$) are, in
 fact, larger than those for the OC pseudospinors. The same problem take place
 when the partial potentials are generated for the V (pseudo)spinors and then
 applied for the OC shells with the same ($lj$). Therefore, in general (see
 \cite{GRThGr} for more theoretical details), the Hamiltonian matrix elements
 are not so accurately reproduced as in the case of GRECPs. By other words,
 the radius of ``unphysical'' terms in a conventional RECP is substantially
 larger than the matching radii (contrary to the GRECP case as one can see
 from our papers and the aforegoing text).

 In turn, the ECAPP generation schemes take no care about smooth
 pseudoorbitals (pseudospinors) and matching radii when putting simulation of
 some valence energies and generation of smooth partial potentials on the first
 place. (A small number of gaussians in the ECAPP expansion is, in fact,
 equivalent to smooth partial potentials). The result is similar to that in
 the case of the conventional shape-consistent RECP. Additional disadvantages
 of ECAPPs are a poor theoretical justification and technical complexity in
 fitting a large number of valence energies. When taking account of
 correlations in ECPP calculations, the valence energies for the correlated
 states should be fitted as well if a high computational accuracy is required.
 It should be done as a pay for the absence of matching radii for the ECPP
 pseudospinors. Moreover, being {\it one-electron} potentials of a special
 (radially local) type, ECAPPs generated for a fixed number of explicitly
 treated electrons can not provide arbitrarily high accuracy even for
 reproducing the one-electron part of the valence Hamiltonian\ \cite{tav00Ab},
 not to mention the {\it two-electron} part. Besides, how many
 transition/valence energies between/for correlated states should be fitted for
 reliable reproducing a required (large) number of the two-electron integrals
 with some needed accuracy? Hundreds? Or maybe thousands? Can it be efficiently
 applied in practice? Will it provide a ``proportional'' (even) level of
 errors for the one- and two-electron integrals as is in the case of GRECPs?
 Should not forget that the ECAPPs employ the conventional RECP operator that
 is not so flexible as the GRECP one\ \cite{GRThGr,tav00Ab}. We can also remind
 here about those properties which cannot be calculated from potential curves
 or surfaces. Why will these properties be accurately reproduced with the help
 of ECAPPs?

\vspace{0.3cm}
 In the optimization of the parameters of partial potentials $V_{nlj}(r)$ one
 can produce compact gaussian expansions for the ECAPPs when fitting directly
 some selected valence energies. Is this a real advantage? The compactness in
 the gaussian expansions of the partial potentials does not ensures the
 smoothness of pseudoorbitals (pseudospinors). Moreover, the radius of
 unphysical terms in such a PP is invitably larger than in the RECPs in which a
 large set of gaussians is employed to fit quite singular behaviour of
 numerical potentials $(V_{nlj}(r){-}N_{core}/r)$ close to the matching points
 thus reducing their effective radii (see Figure~2 in\ \cite{GRThGr}). It is
 very widely known that the effort in the calculation ($\sim N^4$) and
 transformation ($\sim N^5$) of two-electron integrals (where $N$ is the number
 of basis functions) is {\it always} substantially higher than in calculation
 of the RECP integrals ($\sim N^2 \cdot N_{RECP}$, where $N_{RECP}$ is the
 number of terms in the used RECP expansions) for all the known RECP versions
 including GRECPs when appropriately large basis sets are employed for precise
 calculations.
 Again we should emphasize that in spite of the rather complicated form of the
 GRECP operator, the main computational effort in calculating matrix elements
 with GRECPs is caused by the standard radially local operator which is also
 the main part of the shape-consistent RECP and ECAPP operators, and not by the
 non-local GRECP terms. Thus, the additional efforts in calculations with
 GRECPs are negligible as compared with the cases of using the conventional
 RECPs and PPs if comparable gaussian expansions are used for the partial
 potentials.

 Summarizing, what is the computational utility in generation of compact RECPs
 and PPs? Maybe much more important is that their accuracy should be in
 agreement with the number of explicitly treated electrons? Obviously, the
 smooth shapes of pseudospinors in the atomic core are more important than
 smooth partial potentials. The smooth pseudospinors can be accurately
 approximated with a relatively small number of gaussian functions.
 A possibility to generate the partial potentials after constructing
 pseudospinors (and not simultaneously) when inverting the Hartree-Fock
 equations following Goddard~III (1968)\ \cite{Goddard} is very convenient from
 the computational viewpoint. Due to these features, the RECP generation scheme
 by K.Pitzer's group is very effective in practice. It is always better to
 have a possibility to split solution of a complicated problem on a few
 consequent steps. That is the reason why we prefer the scheme of K.Pitzer's
 group and not that proposed by Durand \& Barthelat (1975) in their classical
 paper\ \cite{Durand}, where the idea of the shape-consistent ECP method was
 first suggested. The ECAPP generation scheme is in many aspects close to the
 ECP generation scheme by Durand \& Barthelat.

\vspace{0.3cm}
 Our first comparisons of RECPs in\ \cite{Mos97} were done in the
 one-configurational approximation because only recently we have obtained an
 opportunity of employing very efficient atomic Relativistic Coupled Cluster
 (RCC)\ \cite{Kaldor} code for reliable correlation structure calculations with
 both Dirac-Coulomb and (G)RECP Hamiltonians\ \cite{Mos00Hg}. The advantage of
 the former comparison is in the use of the finite-difference method (i.e.\
 spinors (orbitals) are varied in the numerical form) and therefore, the DF
 (HF) calculations are independent of the finite basis sets errors.
 In the latter case, one has a possibility to use very large basis sets thus
 minimizing dependence of the final results on a special choice of a basis set.
 Besides, there are almost no subjective dependences in RCC calculations from a
 special selection of configurations (reference spaces, truncation thresholds,
 etc.) that is very important for correct comparison of different effective
 Hamiltonians with original.
 Our statements in\ \cite{Mos97} concerning the accuracy of the GRECP
 Hamiltonian were done on the bases of one-configurational calculations in the
 $jj$-coupling scheme and of the theoretical analysis presented in\
 \cite{GRThGr}. They are completely confirmed in first correlation calculations
 of Hg\ \cite{Mos00Hg}, Pb\ \cite{tim00} and TlH\ \cite{tav00TlH}. Besides, the
 examined energy-adjusted PPs were found in our tests to be less accurate in
 general than the shape-consistent RECP versions generated by other groups.


\section{Reply on remarks of M.~Dolg}

 Below Dolg's remarks from the discussed paper and our answers are given. All
 Dolg's quotations are taken from section 5.3 ``Limitations of accuracy'' in
 the same order as in the discussed paper unless the opposite is explicitly
 stated.

 To minimize problems with ambiguous treatment of the quotations when
 extracting them from a context, the whole text of section 5.3 is presented in
 Appendix\ \ref{AppendA}.

\begin{enumerate}

\item      See Table 1 in the Dolg's paper.\\

\item[*]   {\it Our remarks:}\\
 The frozen core approximation is underlying for all the known ECP methods,
 both nonrelativistic and relativistic. Therefore, the accuracy of the (R)ECPs
 can not be considered as higher than that of the frozen core approximation
 unless some special corrections like the Core Polarization Potential (CPP) or
 our Self-Consistent (SfC) terms are used. Moreover, the smoothing of the
 orbitals (spinors), incorporating the relativistic effects, etc.\ will further
 increase the RECP errors. In Table\ \ref{FCA}, we have compiled the errors of
 the new ECPP with 54 adjustable parameters from Table 1 in the Dolg's paper
 together with the frozen core approximation errors calculated by us. The HFD
 code\ \cite{HFD} is used in the corresponding Dirac-Fock all-electron and
 frozen core calculations with the point nuclear model for the states averaged
 over the nonrelativistic configurations. Obviously, having 54 adjustable
 parameters in the ECPP, one can use them to fit exactly 54 valence energies.
 However, the accuracy of the generated PP should not be estimated by the
 errors in reproducing the fitted energies. The valence energies which were not
 used in the fitting procedure or other properties must be used for the PP
 testing because the basic requirement of any fruitful simulation is a
 transferability of a model Hamiltonian to the cases which were not used when
 constructing this Hamiltonian.
 Although the number of states which is of interest for electronic structure
 calculations of an atom is usually not too large to allow one to fit all of
 them, this number is dramatically increased in the case of its chemical
 bounding with a multitude of other atoms.


 Therefore, the accuracy of the new ECPPs can not be derived from Tables 1 and
 2 in the Dolg's paper and additional independent testing is necessary.
 Unfortunately, we can not do this because we do not have the parameters of his
 new ECPPs (see\ \cite{refuse}).


\item      Dolg:\\
 {\it Tables 1 and 2 demonstrate that for very special cases like Hg, with a
 closed 5d$^{10}$-shell in all electronic states considered, a small-core
 energy-consistent pseudopotential using a semilocal ansatz reaches an accuracy
 of 10 cm$^{-1}$, which is well below the effects of the nuclear model, the
 Breit interaction or higher-order quantum electrodynamical contributions. We
 also note that differences between results obtained with a frequency-dependent
 Breit term and the corresponding low-frequency limit amount to up to 10
 cm$^{-1}$. Moreover, the quantum electrodynamic corrections listed in tables
 1 and 2 might change by up to 20 cm$^{-1}$ when more recent methods of their
 estimation are applied\ \cite{Pyykko,Labzowski}}
 \footnote{Papers within the quotations are cited according the list of
 references in the present comments and the numbers of tables are original.}\\

\item[*]   {\it Answer:}\\

 The first part of the quotation is commented in the previous item and in
 Section\ \ref{Comparison}.
 Some special remarks can be made with respect to the Breit effect. The
 replacement of the two-electron Coulomb-Breit interaction by the two-electron
 Coulomb interaction ($1/r_{12}$) plus one-electron PP operator in the PP
 Hamiltonian is not justified by M.~Dolg.
 The contribution from the (frequency-dependent) Breit interaction was
 evaluated in the first-order perturbation theory (PT1). However, the Breit
 interaction is very strong close to a heavy nucleus. Therefore, the wave
 function in its neighborhood is seriously perturbed by the Breit interaction
 and the higher PT orders by correlation should be considered (to describe the
 core relaxation) for appropriate accounting for the Breit correction\
 \cite{Lindroth}. In particular, the random phase approximation can be used
 keeping only the first-order perturbation on the Breit interaction, in which
 it is correct when accounting for the quantum electrodynamic (QED) effects.
 As is shown in\ \cite{Lindroth,Kozlov00}, the core relaxation can reduce the
 final Breit correction by an order of magnitude. Did Prof.\ Dolg perform
 similar analysis when generating the ECPPs for Hg? After that, what is the
 need to take account of the QED effects in the framework of the ECPP when they
 give an order of magnitude smaller contribution than the
 ECPP errors arising from the radially local form of the ECPP operator? What
 is the profit (advantage) in such an accounting for the Breit and QED effects?

 Those ``improvements'' are done by Prof.\ Dolg in the ECAPP method which can
 be done easily and not those which should be done first of all.
 When developing the GRECP method, we are eliminating at first its largest
 errors, then errors of the next level of magnitude and so on, step-by-step.
 The theoretical analysis of the GRECP errors is always done, thus justifying
 the approximations made by us.

 At last, it would be excellent to perform molecular calculations on a level
 of accuracy of 100 cm$^{-1}$ for transition, dissociation, etc.\ energies
 systematically but the modern correlation methods, codes and computers do not
 allow one to do this because of the high computational cost. Our goal on the
 nearest future is to generate the GRECPs with inherent errors close to (or
 below than) 100 cm$^{-1}$ for the valence energies when treating minimal
 number of electrons explicitly. In molecular GRECP calculations, it allows
 one to attain accuracy within a few hundred wave numbers for the energies of
 interest reliably and with minimal efforts.

\item      Dolg:\\
 {\it ``Therefore, it is important to state exactly which relativistic
 all-electron model the effective core potential simulates and, when comparing
 effective core potentials of different origins, to separate differences in the
 underlying all-electron approach from errors in the potential itself, e.g.,
 due to the size of the core, the method of adjustment or the form of the
 valence model Hamiltonian.''}

\item[*]   {\it Answer:}\\
 It is true. 
 As one can see from our papers, we are carefully analyzing the sources of
 errors in our GRECP versions. We {\it ``state exactly which relativistic
 all-electron model the effective core potential simulates''}, etc.
 Moreover, we consider as our duty to present all the necessary details
 concerning all the GRECPs which were used in our papers. Being requested, the
 GRECP parameters can be received, in particular, by email.

 However, it is not in our responsibility {\it ``to separate differences in
 the underlying all-electron approach from errors in the potential itself''}
 for PPs and RECPs generated by other groups. How can we separate errors of the
 Wood-Boring approximation from the ECAPP fitting errors without knowledge
 of all the details of fitting, without having the required codes, and without
 doing some test calculations with these codes? Besides, why {\it must} we do
 this? The responsibility for such an analysis is on those who have generated
 these PPs and RECPs.

 We have written in \cite{Mos00Hg}:\\ 
 ``It should be noted that the energy-adjusted pseudopotential (PP) tested in
 the present paper was generated by H\"aussermann {\it et al.}\ \cite{Hausser}
 using the results of the quasirelativistic Wood-Boring \cite{Wood} SCF
 all-electron calculations as the reference data for fitting the spin-orbit
 averaged PP parameters. A new 20e-PP for Hg was generated recently by fitting
 to the Dirac-Fock-Breit reference data\ \cite{Dolg}, but we do not have the
 parameters of this PP\footnote{See Ref.\ \cite{refuse} for more details.}.

 The energies of transitions between the $6s^2$ and $6s^1 6p^1 (^3P_0, ^3P_1,
 ^3P_2)$ states in the 20e-PP/MRCI calculations employing the CIPSO method\
 \cite{CIPSO} are within 100~cm$^{-1}$ of experiment (see Table~6 in\
 \cite{Hausser} or Table~2 in the present paper). However, the energy-adjusted
 PP does not account for the contributions from correlations with the $4f$
 shell, and the basis set used does not contain $h$-type functions. One can see
 from Tables 1 and 2 that the contributions of the two effects to these
 transition energies are up to 284~cm$^{-1}$ and 247~cm$^{-1}$, respectively.
 The good 20e-PP/MRCI/CIPSO results are probably due to fortuitous cancellation
 of several contributions: the inherent PP errors (e.g., the
 $6s^1_{1/2}6p^1_{1/2}(J=0)$ -- $6s^1_{1/2}6p^1_{3/2}(J=2)$ splitting is
 overestimated by 1014 cm$^{-1}$ because of the features of the spin-orbit
 simulation within the $LS$-based version of the energy-adjusted scheme, see
 Table~4 in paper~\cite{Mos97}), the neglect of correlations with the $4f$
 shell, the basis set incompleteness, etc. A similar situation holds for
 transitions between the $6s^1$ and $6p^1 (^2\!P_{1/2}, ^2\!P_{3/2})$ states of
 Hg$^+$, but errors of the 20e-PP/MRCI/CIPSO calculations relative to
 experimental data reach a level of 1000~cm$^{-1}$ in this case.''

 Is it not correct? Similar analysis can be found in our previous paper\
 \cite{Mos97} criticized by M.~Dolg.
 M.~Dolg many times claimed that our test results with their 20e-PP in\
 \cite{Hausser} are wrong. Where can we find his publication with the
 confirmation of these claims?

 However, let us get back to the paper of M.~Dolg.
 Why is the information about the states used in the valence energy fitting in
 the ECPP generation not even presented there? Where are the ECPP parameters?
 As is mentioned above, we have no a possibility to check the real quality of
 these new ECPPs (see\ \cite{refuse}).


\item      Dolg:\\
 {\it ``In this context we want to point out that the seemingly large errors
 for energy-adjusted pseudopotentials reported by Mosyagin et al.\
 \cite{Mos97,Mos00Hg} are mainly due to the invalid comparison of
 Wood-Boring-energy-adjusted and Dirac-Fock-orbital-adjusted pseudopotentials
 to all-electron Dirac-Fock data, i.e., differences in the all-electron model
 are considered to be pseudopotential errors.''}

\item[*]   {\it Answer:}\\
 Although ``{\it the correct relativistic all-electron Hamiltonian for a
 many-electron system is not known}'', the Dirac-Coulomb Hamiltonian is
 preferred over the Wood-Boring one. Moreover, for an ``RECP user'' the level
 of the PP errors with respect to the most accurate relativistic Hamiltonian
 (among the known ones) is much more meaningful than the question whether the
 PP errors are due to the unsatisfactorily fitting procedure of the ECAPPs (the
 small number of parameters, incompleteness of the PP operator, etc.) or due
 to the poor all-electron reference data used for this fitting. Therefore, the
 comparison of the all-electron Dirac-Coulomb data with the Wood-Boring-fitted
 PP results is correct in papers\ \cite{Mos97,Mos00Hg}, whereas the manner of
 comparison of the all-electron Wood-Boring data with the Dirac-Fock-based
 GRECP results in Table XVII from\ \cite{Mos97} by M.~Dolg is not valid. The
 (G)RECP and PP results are given in\ \cite{Mos97} {\it only} in order to show
 ``the range of the dispersion of the data''. Besides the absence of
 equivalence in the used basis sets, the Wood-Boring approximation is an
 additional source for the distinctions between the all-electron and GRECP
 molecular data in this table (see also the last item in this section for more
 details). The question is also arise how M.~Dolg and co-authours can attain
 an ``{\it excellent agreement}'' (see abstract of \cite{Hausser}) with the
 experimental data in their previous papers with the help of the Wood-Boring
 fitted PPs.




\item      Dolg:\\
 {\it ``It is also obvious from the compiled data that the accuracy of the
 valence model Hamiltonian is also a question of the number of adjustable
 parameters.''}

\item[*]   {\it Answer:}\\
 We are very satisfied that Prof.\ Dolg at last have recognized the fact of
 importance of the number of the adjustable parameters because probably all the
 ECAPPs generated before had small number of the parameters. The problem is
 only that the ECAPP is a one-electron operator and the original Hamiltonian
 contains the two-electron interactions as well. How is Prof.\ Dolg planning to
 reproduce the two-electron part when fitting the ECAPP parameters?

\item      Dolg:\\
 {\it `` Claims that such very high accuracy as demonstrated here can only be
 achieved by adding nonlocal terms for outer core orbitals to the usual
 semilocal terms \cite{Mos97,Mos00Hg} appear to be invalid, at least for
 energy-consistent pseudopotentials.''}

\item[*]   {\it Answer:}\\
 We have revised again our papers but could not find the text which could be
 interpreted by such a manner as it is done in the Dolg's paper. Can Prof.\
 Dolg show the places in our papers where we have written so? The most
 debatable phrase from our papers which can be associated with the above Dolg's
 quatation is 
 ``...The larger errors for RECPs\ \cite{Ross,Hausser} are mainly due to the
 neglect of the difference between the outer core and valence potentials in
 these RECP versions (see\ \cite{GRThGr,Tup95} for details).''
 We have written it on p.~674 of our joint paper\ \cite{Mos00Hg} with the
 Tel~Aviv group but this phrase is in the responsibility of the authors of the
 present comments. However, ``larger errors'' means here ``larger level of
 errors'' (it is clear from the context below this phrase where we clarify the
 origin of the errors for our particular calculations and for the used 20e-PP
 for Hg).  Nevertheless, we are ready to recognize that in the case of the
 20e-PP for Hg\ \cite{Hausser} the main contribution to the errors in the
 transition energies compiled in Table 3 of\ \cite{Mos00Hg} is due to a bad
 quality of the used fitting principles and/or incompetent their application
 in\ \cite{Hausser} rather than due to the neglect of the difference between
 the outer core and valence potentials in this PP, if such a reformulation is
 more acceptable for authors of\ \cite{Hausser}.

 Obviously, some errors in transition energies can be smaller for an ECAPP,
 especially, if these energies are fitted when generating the ECAPP. We have
 pointed out earlier, that the ECAPPs should be checked for those transitions
 or properties, which were not fitted during the ECAPP generation. Test
 calculations should be performed with different numbers of correlated
 electrons and with a good quality of accounting for correlation. A ``minimal
 completeness'' of the basis sets is also required.

 As to the importance of the nonlocal terms and to the phrase
 {\it ``This error could be reduced further upon using a smaller core, but the
 efficiency of the approach would be sacrificed.''} written by Prof.\ Dolg a
 few lines above, we should remind the following. 
 Already in paper\ \cite{Tup95} we have emphasized that when freezing the OC
 pseudospinors, the corresponding nonlocal GRECP projectors are not involved in
 calculations. However, the accuracy can be very high if partial potentials are
 generated for {\it nodal} V pseudospinors and not for OC ones thus taking into
 account the difference between the V and OC potentials contrary to the
 standard RECP case. Therefore we considered this frozen core scheme as a
 special GRECP version. We have pointed out later (e.g., see\ \cite{GRThGr})
 about similar alternatives with respect to other our additions
 (``self-consistent'' and ``spin-orbit'' terms) to the conventional radially
 local operator.
 Obviously, the same (freezing) procedure can be applied when more core
 shells are included into the space of explicitly treated electrons. Freezing
 then the innermost shells of them, one again can remove the nonlocal
 GRECP terms because the differences between the partial potentials for the
 innermost nodeless pseudospinors and the next pseudospinors with the same
 ($lj$) (but having one node) are the most essential to be taken into account.
 However these cases are not computationally interesting because such RECP
 are of interest for the modern correlation structure calculations which can
 provide a required accuracy when treating as small number of electrons
 explicitly as possible. The latter is our {\it main} purpose. That is why we
 prefer to change the functional form of the RECP operator, to insert core
 correlations to GRECPs etc.
 The only problem is to do this properly, by a ``theoretically-consistent''
 way, involving appropriate functional forms.  And this should be done when
 solving step-by-step the most actual problems, i.e.\ eliminating first the
 sources of the largest errors with minimal complication in the resulting RECP
 calculations.  Accounting for the (high PT order) QED and other effects within
 an RECP is meaningless if they have smaller order of magnitude than inherent
 errors of the RECP under consideration.

\item      Dolg:\\
 {\it ``Moreover, additional nonlocal terms obviously do not improve
 the performance for atomic states with a $5d^9$ occupation''}

\item[*]   {\it Answer:}\\
 It is not true. M.~Dolg did not present {\it any} his results of calculations
 with GRECPs or {\it any} theoretical analysis. As we know from our
 correspondence with him, similar work have not been performed by him at all.
 Therefore, his conclusions are made only on the basis of our results given in
 \cite{Mos97}. However, our results and conclusions in\ \cite{Mos97} are
 opposite. In this connection, the adverb {\it ``obviously''} is very funny.

\item      Dolg:\\
 {\it ``or in molecular calculations (cf., e.g., tables III and XVII in
 Mosyagin et al. \cite{Mos97}).''}

\item[*]   {\it Answer:}\\
 Concerning the molecular GRECP calculations, the first GRECP/MRD-CI results
 for spectroscopic constants in TlH\ \cite{tav00TlH} again lead to opposite
 conclusions. Some other GRECP/MRD-CI and GRECP/RCC-SD calculations on HgH, TlH
 and PbH are in progress now. As to the spin-orbit-averaged (G)RECP/SCF and
 PP/SCF results on HgH presented in table XVII of\ \cite{Mos97}, we are forced
 to repeat that our HgH calculations were performed there in order to study
 ``the range of dispersion of the data'' because for the {\it one-electron}
 RECP and GRECP operators it ``is important information to estimate the
 accuracy of the RECP approximation both for one-configurational and for highly
 correlated calculations of HgH and HgH$^+$ molecules'' (see p.~1121 in\
 \cite{Mos97}).
 The above mentioned Dolg's excerption is certainly the top analytical {\it
 result} in the commented paper dealing with the {\it pseudo}potentials.
 Following its logic pattern, we can call this by a {\it pseudoresult} (and
 moreover, the {\it results} obtained with {\it effective} potentials can be
 analogously called by {\it effective results} for ``sake of brevity'').

\end{enumerate}

 In fact, almost all that we have written in these comments was written in our
 papers earlier and we only have concentrated here on some underlying
 principles of our approach as compared to other RECP methods. We regret that
 our papers have occured to be so difficult for reading that Prof.\ Dolg could
 not clarify the principles and features of the GRECP method.  

 We should add that some more remarks could be given concerning the
 application by Stuttgart group of the core polarization potential together
 with ECAPPs,
 their ``{\it idea to fit exclusively to quantum mechanical observables like
 total valence energies}'' (see the discussed paper), the features of the ECAPP
 operator,
 actions to avoid admixture of the inner core states which are occupied by the
 electrons eliminated from calculations etc. Are the valence energies obtained
 from the Dirac-Fock-Breit equations observable?
 Without a good theoretical justification of the transferability and proper
 application of these very progressive ideas to other problems, the result can
 be unsatisfactory. This we have seen on example of application of the
 Wood-Boring aproximation to the ECAPP generation. Moreover, when developing a
 new method, one should at least to take into account the basic achievements in
 this field made earlier. Besides, the accuracy and reliability of a newly
 developed method should not be lower than that of already existing methods if
 their application require the same computational efforts. Therefore we
 consider the present ECAPP technique as a step in back direction as compared
 to the shape-consistent RECP version developed more than twenty years ago.

 At last we should say that
 we did not find any serious scientific analysis of our conclusions and
 results in the Dolg's paper but only some ``political declarations'' are
 there. Therefore, we are not going to answer in future on claims of similar
 quality as in the commented paper only because do not want to lose time on
 such a level of discussion as is proposed by Prof.\ Dolg.
 Any well-justified critical remarks concerning our GRECPs (or the text in our
 papers) are welcomed. We will answer with pleasure on questions dealing with
 RECPs. We are ready to (and welcome) any public discussion on the RECP methods
 (e.g., within the REHE newsletters) if they will be of common interest.

\acknowledgments
 We are grateful to M.~Dolg for sending us the discussed paper that have
 stimulated us for writing these comments.

 The work on development of the GRECP method was supported by the DFG/RFBR
 grant N 96--03--00069, the INTAS grant No 96--1266 and by the RFBR grant N
 99--03--33249.


\begin{table}[h]
 \caption{ The frozen core approximation (FCA) errors calculated by us with
 the help of the HFD\ \protect\cite{HFD} code at the all-electron Dirac-Fock
 level within the point nuclear model for the states averaged over the
 nonrelativistic configurations. The errors of the new ECPP from Table 1 in
 the Dolg's paper.  All values are in cm$^{-1}$. }
 \label{FCA}
\vspace{5mm}
 \begin{tabular}{llrr}
 \multicolumn{2}{c}{configuration}&\multicolumn{2}{c}{error}\\
\cline{3-4}
           &            &FCA$^a$&ECPP$^b$\\
 \hline                                 
 Hg        & $6s^2$     &  0.0   &  0.0  \\
           & $6s^1 6p^1$&  0.6   &  0.0  \\
 Hg$^+$    & $6s^1$     &  1.0   &  0.0  \\
           & $7s^1$     &  4.0   &  0.0  \\
           & $8s^1$     &  4.3   &  0.1  \\
           & $9s^1$     &  4.4   & -0.1  \\
           & $6p^1$     &  3.3   &  0.0  \\
           & $7p^1$     &  4.3   &  0.0  \\
           & $8p^1$     &  4.4   &  0.0  \\
           & $9p^1$     &  4.5   &  0.0  \\
 Hg$^{++}$ &            &  4.5   &  0.0  
 \end{tabular}
\vspace{3mm}
 $^a$~frozen core approximation with the $1s,\ldots,4f$ frozen shells
 taken from the $6s^2$ Hg state.

 $^b$~energy-consistent pseudopotential with 54 adjustable parameters.

\end{table}


\appendix
 \section{}
  \label{AppendA}

 Because the reader can be not familiar with the criticized paper of Prof.\
 Dolg, below we present its section 5.3, which is discussed in our comments,
 without any changes.

 \subsection*{M.\ Dolg, Section 5.3  ``Limitations of accuracy''}

 Effective core potentials are usually derived for atomic systems at the
 finite difference level and used in subsequent molecular calculations using
 finite basis sets. They are designated to model the more accurate all-electron
 calculations at low cost, but without significant loss of accuracy.
 Unfortunately the correct relativistic all-electron Hamiltonian for a
 many-electron system is not known and the various pseudopotentials merely
 model the existing approximate formulations. For most cases of chemical
 interest, e.g., geometries and binding energies, it usually does not matter
 which particular Hamiltonian model is used, i.e., typically errors due to the
 finite basis set expansion or the limited correlation treatment are much
 larger than the small differences between the various all-electron models.  

\begin{table}[h]
 Table 1. Relative average energy of a configuration of Hg from all-electron
 (AE) multi- configuration Dirac-Hartree-Fock (DHF) average level calculations
 using the Dirac-Coulomb (DC) Hamiltonian with a finite nucleus with Fermi
 charge distribution (fn) or a point nucleus (pn). Contributions from the
 frequency-dependent Breit (B) interaction (frequency of the exchanged photon
 10$^3$ cm$^{-1}$) and estimated contributions from quantum electrodynamics
 (QED, i.e., self-interaction and vacuum polarization) were evaluated in
 first-order perturbation theory. Errors of energy-consistent pseudopotentials
 (PP) with 20 valence electrons and different numbers of adjustable parameters
 with respect to the AE DHF(DC,pn)+B+QED data. All values in cm$^{-1}$.
 \begin{tabular}{llrrrrrr}
 \multicolumn{2}{c}{configuration}&\multicolumn{2}{c}{AE, DHF}&\multicolumn{2}{c}{contribution}&\multicolumn{2}{c}{error}\\
           &             &\multicolumn{2}{c}{(DC)+B+QED}& &     &      &      \\
           &             &  fn      & pn       & B      & QED   &PP$^a$&PP$^b$\\
 \hline
 Hg        & $6s^2$      &        0 &        0 &    0.0 &   0.0 &  0.0 &  0.0 \\
           & $6s^1 6p^1$ &  35632.3 &  35674.4 &  -52.5 & -18.7 &  1.3 &  0.0 \\
 Hg$^+$    & $6s^1$      &  68842.1 &  68885.1 &  -98.6 & -11.6 & -0.1 &  0.0 \\
           & $7s^1$      & 154127.4 & 154206.2 & -220.6 & -42.4 & -0.4 &  0.0 \\
           & $8s^1$      & 178127.5 & 178215.5 & -238.4 & -41.7 &  1.1 &  0.1 \\
           & $9s^1$      & 188751.0 & 188843.2 & -244.1 & -40.6 &  1.6 & -0.1 \\
           & $6p^1$      & 122036.8 & 122128.9 & -154.2 & -41.8 &  0.6 &  0.0 \\
           & $7p^1$      & 167514.3 & 167609.2 & -224.1 & -40.3 & -3.3 &  0.0 \\
           & $8p^1$      & 183808.0 & 183903.6 & -238.5 & -40.0 & -0.8 &  0.0 \\
           & $9p^1$      & 191697.2 & 191793.1 & -244.0 & -39.6 &  0.6 &  0.0 \\
 Hg$^{++}$ &             & 206962.2 & 207058.4 & -249.8 & -39.5 &  2.6 &  0.0 \\
 \end{tabular}
 $^a$ energy-consistent pseudopotential with 26 adjustable parameters. 
 
 $^b$ energy-consistent pseudopotential with 54 adjustable parameters.
\end{table}

 For very accurate calculations of excitation energies, ionization potentials
 and electron affinities, or for a detailed investigation of errors inherent in
 the effective core potential approach, however, such differences might become
 important. Tables 1 and 2 demonstrate that for very special cases like Hg,
 with a closed 5d$^{10}$-shell in all electronic states considered, a
 small-core energy-consistent pseudopotential using a semilocal ansatz reaches
 an accuracy of 10 cm$^{-1}$, which is well below the effects of the nuclear
 model, the Breit interaction or higher-order quantum electrodynamical
 contributions. We also note that differences between results obtained with a
 frequency-dependent Breit term and the corresponding low-frequency limit
 amount to up to 10 cm$^{-1}$. Moreover, the quantum electrodynamic corrections
 listed in tables 1 and 2 might change by up to 20 cm$^{-1}$ when more recent
 methods of their estimation are applied
 \footnote[98]{P.~Pyykk\"o, M.~Tokman, and L.N.~Labzowski, Estimated
 valence-level Lamb shifts for group 1 and group 11 metal atoms, Phys.\ Rev.\ A
 {\bf 57}, R689 (1998).}$^,$\footnote[99]{L.~Labzowski, I.~Goidenko, M.~Tokman,
 and P.~Pyykk\"o, Calculated self-energy contributions for an ns valence
 electron using the multiple-commutator method, Phys.\ Rev.\ A {\bf 59}, 2707
 (1999).}.
 Therefore, it is important to state exactly which relativistic all-electron
 model the effective core potential simulates and, when comparing effective
 core potentials of different origins, to separate differences in the
 underlying all-electron approach from errors in the potential itself, e.g.,
 due to the size of the core, the method of adjustment or the form of the
 valence model Hamiltonian.  In this context we want to point out that the
 seemingly large errors for energy-adjusted pseudopotentials reported by
 Mosyagin et
 al.\footnote[100]{N.S.~Mosyagin, A.V.~Titov, and Z.~Latajka, Generalized
 Relativistic Effective Core Potential: Gaussian expansions of potentials and
 pseudospinors for atoms Hg through Rn, Int.\ J.\ Quant.\ Chem.\ {\bf 63}, 1107
 (1997).}$^,$\footnote[101]{N.S.~Mosyagin, E.~Eliav, A.V.~Titov, and U.~Kaldor,
 Comparison of relativistic effective core potential and all-electron
 Dirac-Coulomb calculations of mercury transition energies by the relativistic
 coupled-cluster method, J.\ Phys.\ B {\bf 33}, 667 (2000).}
 are mainly due to the invalid comparison of Wood-Boring-energy-adjusted and
 Dirac-Fock-orbital-adjusted pseudopotentials to all-electron Dirac-Fock data,
 i.e., differences in the all-electron model are considered to be
 pseudopotential errors.

 Note that in the above example of Hg the average energy of a configuration
 (table 1) and the fine-structure (table 2) of one-valence electron states is
 more accurately represented than the fine-structure of the $6s^1 6p^1$
 configuration. The small errors in the latter case are a consequence of the
 pseudoorbital transformation and the overestimation of the 6s-6p exchange
 integral with pseudo-valence spinors. This error could be reduced further upon
 using a smaller core, but the efficiency of the approach would be sacrificed.
 It is also obvious from the compiled data that the accuracy of the valence
 model Hamiltonian is also a question of the number of adjustable parameters.
 Claims that such very high accuracy as demonstrated here can only be achieved
 by adding nonlocal terms for outer core orbitals to the usual semilocal
 terms$^{100,101}$ appear to be invalid, at least for energy-consistent
 pseudopotentials. Moreover, additional nonlocal terms obviously do not improve
 the performance for atomic states with a $5d^9$ occupation or in molecular
 calculations (cf., e.g., tables III and XVII in Mosyagin et al.$^{100}$).

\begin{table}[h]
 Table 2. As table 1, but for fine-structure splittings. All values in
          cm$^{-1}$.
 \begin{tabular}{lllrrrrrr}
 \multicolumn{2}{c}{configuration}& splitting &\multicolumn{2}{c}{AE,DHF}&\multicolumn{2}{c}{contribution}&\multicolumn{2}{c}{error}\\
        &             &                       &\multicolumn{2}{c}{(DC)+B+QED}& &     &      &      \\
        &             &                       &  fn     & pn      & B      & QED &PP$^a$&PP$^b$\\
 \hline
 Hg     & $6s^1 6p^1$ & $^3P_1-^3P_0$         &  1987.7 &  1988.6 &  -25.5 & 0.9 & -14.7&  3.0 \\
        &             & $^3P_3-^3P_0$         &  6082.6 &  6084.8 &  -96.8 & 2.9 & -28.3& -3.5 \\
        &             & $^1P_1-^3P_0$         & 22994.4 & 22982.3 &  -72.4 & 2.2 & -12.4& -9.4 \\
 Hg$^+$ & $6p^1$      & $^2P_{3/2}-^2P_{1/2}$ &  7765.3 &  7768.8 & -132.8 & 4.8 & -14.8& -0.1 \\
        & $7p^1$      & $^2P_{3/2}-^2P_{1/2}$ &  2136.8 &  2137.9 &  -29.0 & 1.1 &  -1.7&  0.2 \\
        & $8p^1$      & $^2P_{3/2}-^2P_{1/2}$ &   939.4 &   939.9 &  -12.1 & 0.4 &  -4.6& -0.3 \\
        & $9p^1$      & $^2P_{3/2}-^2P_{1/2}$ &   498.7 &   498.9 &   -6.2 & 0.2 &  -3.5&  0.0 \\
 \end{tabular}
 $^a$ energy-consistent pseudopotential with 26 adjustable parameters. 
 
 $^b$ energy-consistent pseudopotential with 54 adjustable parameters.
\end{table}


\begin{references}
\bibitem{refuse}      Prof.\ Dolg twice refused to send us the parameters of
                his new ECPPs for Hg, at first, when we were preparing paper\
                \cite{Mos00Hg} (and therefore we were forced to use their
                previous PP version for Hg in our RECP comparison in\
                \cite{Mos00Hg}) and then, after publishing the results with
                the new ECPPs in the commented paper.
\end{references}
\end{document}